\documentclass{elsart}

\usepackage{natbib}

 \usepackage{graphicx}

\usepackage{amssymb}

\begin{document}

\begin{frontmatter}



\title{Spectral Energy Distribution Fitting:  Application to Ly$\alpha$-Emitting Galaxies\thanksref{talk}}

\thanks[talk]{Based on a presentation and discussion at ``Understanding Lyman-alpha Emitters'', MPIA, Heidelberg, Oct. 6-10, 2008}


\author{Eric Gawiser}
\address{Department of Physics and Astronomy, Rutgers University}
\ead{gawiser@physics.rutgers.edu}

\begin{abstract}
Spectral Energy Distribution (SED) fitting is a well-developed 
astrophysical tool that has recently been applied to high-redshift 
Ly$\alpha$-emitting galaxies.  
If rest-frame ultraviolet through near-infrared photometry is available, 
it allows the simultaneous determination of the  
star formation history and dust extinction of a galaxy.  
Ly$\alpha$-emitter SED fitting results from the literature find   
star formation rates  $\sim3$M$_\odot$~yr$^{-1}$, 
stellar masses $\sim10^9$M$_\odot$ for the general population 
but $\sim10^{10}$M$_\odot$ for the subset detected by IRAC, 
and very low dust extinction, $A_V\leq 0.3$, although a couple of 
outlying analyses prefer significantly more dust and higher intrinsic 
star formation rates.  
A checklist of 14 critical choices that must be made when performing 
SED fitting is discussed.
\end{abstract}


\end{frontmatter}

\section{Spectral Energy Distribution Fitting}
\label{sed}
Although it is not widely 
recognized, Spectral Energy Distribution (SED) fitting and photometric redshift determination are identical.  
Both use the likelihood function 
$L(z,T)$ that a galaxy's observed 
SED was generated by template type $T$ at redshift 
$z$.  This is typically generated from 
\begin{equation}
L(z,T) = \frac{ \exp(-\chi^2/2) }{ (2 \pi)^{N/2} \Pi_i \sigma_i} , \; \; \;
\mathrm{where} \; \; \;  \chi^2 = \sum_i (p_i-o_i)^2/\sigma_i^2 , 
\end{equation}
$i$ indexes the 
$N$ photometry bands, $p$ are template predictions in 
these bands, $o$ are observed 
fluxes and $\sigma$ are observational errors.
A best-fit normalization factor for each template/object combination 
can be solved for analytically to simplify optimization.

When only a photometric redshift is desired, one marginalizes or 
optimizes over the nuisance parameter $T$, and it usually suffices to use 
a limited set of templates.  SED fitting typically involves a large set of 
templates at fixed redshift, corresponding to a delta-function prior in $z$.
This is justified when a spectroscopic redshift is available but is a dubious 
practice otherwise.  
A common practice is 
to determine photometric redshifts with a limited template 
set, fix the redshift at the best fit, and then perform SED fitting with a 
wide range of templates.  
This is computationally convenient but statistically inconsistent, 
is guaranteed to underestimate the uncertainties, and runs the risk 
of biasing the results significantly.   

Nonetheless, SED fitting is a robust, well-developed method at low redshift, 
where spectroscopic redshifts are typically available
(see \citealp{kannappang07} 
and references therein for discussions of applications and 
caveats).
 Ly$\alpha$-Emitting (LAE) galaxies are well-suited 
to SED fitting because the narrow-band selection finds galaxies in a narrow 
enough redshift range that a fixed redshift can be assumed
when fitting broad-band photometry.  However,   
the signal-to-noise (S/N) available for photometry of these dim, high-redshift 
galaxies is considerably lower than for other galaxy types to which SED 
fitting has been applied.  Hence we need to 
select SED methods that are appropriate for low S/N and to avoid over-fitting.
Novel methods may be required to handle the 
unusual characteristics of these galaxies, in particular the guarantee from 
high equivalent-width
 emission-line selection that a starburst is occurring at the time of 
observation, the corresponding guarantee that other nebular emission lines 
are strong, and the opportunity to utilize narrow-band photometry as part of 
the SED.  

Despite the low S/N available for LAEs, 
SED fitting has been claimed to allow the determination of star formation rate (SFR),
 stellar mass, stellar 
age, characteristic timescale for star formation ($\tau$), dust extinction, 
and a new $q$ parameter describing radiative transfer effects on 
Ly$\alpha$ photons \citep[e.g.,][]{finkelsteinetal07}.
When a starburst is occurring, the 
burst population is expected to dominate the rest-ultraviolet continuum. 
In the unusual case that dust extinction is negligible, the SFR  is proportional 
to the rest-UV flux \citep{kennicutt98a}
and the rest-UV slope 
depends on 
the age of the starburst and 
the shape of the initial mass function. 
 In most galaxies, dust is too abundant for this interpretation to be 
meaningful, but in LAEs the selection 
method guarantees minimal dust extinction (see Fig. 14 of \citealp{gronwalletal07}).  
If dust extinction is not too high, the rest-NIR luminosity 
density is nearly directly proportional to the stellar mass, 
with some dependence on 
the age of the stellar population traced by rest-optical color 
since older main-sequence 
stars have a higher mass-to-light ratio
\citep{belletal03, portinarietal04}.  

\section{Results for Ly$\alpha$-emitting Galaxies}
\label{results}

Table \ref{tab:results} 
includes all reported results for emission-line-selected 
LAEs as of October, 2008.  This compilation does not include 
objects initially selected in the continuum that turned out 
to show strong Ly$\alpha$ emission \citep{pentericcietal08}; 
these should be referred to 
as, e.g., Ly$\alpha$-emitting Lyman break galaxies, rather than LAEs.  
All observations find similar rest-UV flux densities, hence analyses 
with large inferred dust extinction also report very large intrinsic 
SFR.  
Significant scatter exists in the age results, 
partially due to wide variation in model assumptions
 between constant SFR, $\tau$ and instantaneous burst  
models.  The only robust determinations are of 
star formation rates $\sim3$M$_\odot$~yr$^{-1}$, 
stellar masses $\sim10^9$M$_\odot$ for the general population 
but $\sim10^{10}$M$_\odot$ for the subset detected by IRAC, 
and of very low dust extinction, $A_V\leq 0.3$, although a couple of 
outlying analyses prefer significantly more dust and correspondingly 
higher intrinsic star formation rates.

\begin{table}[!ht]
\caption{Results from 
F09 \citep{finkelsteinetal09}, 
G07 \citep{gawiseretal07}, L07 \citep{laietal07}, L08 \citep{laietal08}, 
N07 \citep{nilssonetal07}, and P07 \citep{pirzkaletal07}.  NI (I) means that 
analysis was restricted to LAEs lacking (having) IRAC detections.
Analyses by G07, L08 and N07  
are of stacked populations, whereas other results show the average and 
scatter amongst results for individual LAEs, with all error bars corresponding 
to the 68\% confidence level.  Where two-population 
fitting was used (G07, F09, P07), 
age and $\tau$ refer to the younger population 
and young fraction is the fraction of the total stellar mass given in the 
5th column that resides in the young population; 
if single-population fitting was used, young fraction is set to 1.  If 
a constant SFR (instantaneous burst) was assumed, $\tau$ is set to $\infty$
 ($0$).  The F09 results for radiative transfer of Ly$\alpha$ photons 
yield $q=1.0^{+2.8}_{-0.4}$.  
}
\label{tab:results}
\smallskip
\begin{center}
{\small
\begin{tabular}{cccccccc}
\hline
\noalign{\smallskip}
Ref&z&SFR&$A_V$&Stellar mass&age&$\tau$&young\\
   & &[M$_\odot$yr$^{-1}$]&[mag]&[$10^9$ M$_\odot$]&[Myr]&[Myr]&fraction\\
\noalign{\smallskip}
\hline
G07\_NI&3.1&$2\pm1$&$0.0^{+0.1}_{-0.0}$&$1.0^{+0.6}_{-0.4}$&$20^{+30}_{-10}$&$750\pm250$&$0.2^{+0.3}_{-0.1}$\\
L08\_NI&3.1&$2\pm1$&$0.0^{+0.3}_{-0.0}$&$0.3^{+0.4}_{-0.2}$&$160^{+140}_{-110}$&$\infty$&$1$\\
L08\_I&3.1&$6\pm1$&$0.0^{+0.3}_{-0.0}$&$9\pm3$&$1600\pm400$&$\infty$&$1$\\
N07&3.15&$0.7^{+0.5}_{-0.3}$&$0.3^{+0.1}_{-0.2}$&$0.5^{+0.4}_{-0.3}$&$850^{+130}_{-420}$&$\infty$&$1$\\
F09&4.5&$140^{+170}_{-110}$&$0.5\pm0.2$&$15^{+35}_{-14}$&$13^{+500}_{-7}$&$0$&$0.4\pm0.4$\\
P07&$\sim$5&$8\pm1$&$0.1\pm0.1$&$0.2^{+0.3}_{-0.1}$&$4^{+4}_{-3}$&$0$&$0.2^{+0.4}_{-0.2}$\\
L07\_I&5.7&$400^{+600}_{-370}$&$0.7\pm0.4$&$17\pm13$&$500\pm400$&$\infty$&$1$\\
\noalign{\smallskip}
\hline
\end{tabular}
}
\end{center}
\end{table}


Discussion about the current results included concerns that systematic effects due to different analysis methods on these parameters can be serious.   
To estimate the severity of this, we agreed to trade SEDs to be analyzed by 
other groups using their methods to see how much difference is produced in the 
best-fit parameters and their reported uncertainties.  
Significant effort should be invested 
in developing a more consistent fitting method than the various ones used to 
obtain the current literature results.  
Determining evolution of LAE results with redshift or comparisons of LAEs 
and LBGs requires applying a common analysis method, as in the posters 
by S. Yuma and K. Ohta at this meeting.

\section{Discussion:  14 SED Fitting Choices}
\label{choices}

In this section we discuss 14 choices necessary for any modeler to make when preparing to fit observed galaxy SEDs.


{\bf 1. Stellar population models:}
One typically chooses a set of stellar population models from 
BC03 (\citealt{bruzualc03}), M05 (\citealt{maraston05}), or the
unpublished CB08 (Charlot-Bruzual 2008) models.  
The M05 and CB08 models include the empirically-determined 
contribution of thermally pulsating asymptotic 
giant branch (TP-AGB) stars, which makes a major 
difference in rest-NIR photometry at intermediate ages.
  So BC03 should be avoided, although the difference is largest at solar metallicity and less severe for the lower metallicities likely for LAEs. 
The discussion pointed out that Starburst99 
\citep{leithereretal99,vazquezl05}
 models include nebular emission, 
which the above models do not.  They have been updated to 
include TP-AGB stars \citep{vazquezl05} and 
should properly handle populations as young as 
1 Myr, whereas the above models should not be trusted for young starbursts due 
to the lack of nebular continuum and emission lines.  However, the effects of metallicity and ionization parameter, both of which are unknown 
for LAEs, become important when using these models.


{\bf 2. Star formation history:}
In hierarchical cosmology, we expect galaxies to have complex star formation 
histories that combine a series of starbursts with quiescent star formation, 
with the average SFR increasing 
versus time during the early stages of galaxy formation before the supply of 
neutral gas becomes exhausted.  In practice, SED-fitting approximates this 
using a smooth history, $SFR(t)=SFR(t_0)\exp(-(t-t_0)/\tau)$, where the SFR is 
assumed to be zero before $t_0$ and the stellar age at the time of 
observation is $t_{obs}-t_0$.  A Constant Star Formation rate (CSF) corresponds to $\tau \rightarrow \infty$ and a Simple Stellar Population (SSP) corresponds 
to a delta-function SFR at $t_0$ (approximated by $\tau \rightarrow 0$). 
 Negative values of $\tau$ should be included 
in fits as they correspond to an exponentially increasing SFR which is 
what would result from a constant {\it specific} star formation rate.  
\citet{finlatoretal07} compared these smooth star formation histories with 
realistic bursty ones from cosmological hydrodynamic simulations and found that 
both produced acceptable fits to high-redshift observations.  

{\bf 3. Minimum age:}
The maximum age that a stellar population should be allowed to have is 
the age of the universe at the time of observation.  This is mildly 
unrealistic, 
as a large SFR at the instant of the Big Bang makes little 
sense, but the observational consequences at $z<6$ 
of beginning star formation at
$z=\infty$ versus e.g., $z=15$ are small given the short time difference.  
The minimum age is a subtler question.  
Following the discussion above, one should use
 Starburst99 (or the equivalent) for modeling any population younger than 
$\sim 10$~Myr while varying the input metallicity and ionization parameter.  

{\bf 4. Initial mass function (IMF):}
Although it is often stated that the IMF appears universal, a  
top-heavy IMF at high-redshift has recently been claimed 
\citep{vandokkum08,dave08}  
and has been used to model LAEs \citep{ledelliouetal06}.  
Even different versions of
 the ``universal'' local IMF make a factor 
of two difference in stellar mass.  This difference can be corrected for 
when comparing results as long as all papers report their assumed IMF.  
During the discussion it was suggested to use a Salpeter IMF between 
100 and 0.01 solar masses as a common standard to enable comparison of 
results.  

{\bf 5. Metallicity range:}
Since the metallicities of LAEs have yet to be measured, a conservative approach is to allow this parameter to vary between 0.01 $Z_\odot$ and $Z_\odot$.  
In theory, SED fitting can be used to fit the metallicity, but the effect is 
minor and is degenerate with other parameters, so it is best to consider this 
as a systematic uncertainty.  The discussion pointed out that assuming a single 
metallicity causes an underestimate in the age uncertainty due to age-metallicity degeneracy in SED shape.  

{\bf 6. Dust law:}
The \citet{calzettietal00} dust law gives the dust extinction as a function of 
wavelength, parametrized by E(B-V)$=0.3$~$A_V$.  Calibrated to local starburst galaxies, this is the dust law typically used for high-redshift SED fitting.
However, we do not know if the dust law evolves with redshift, and given local 
variations we should consider this a significant uncertainty.  It would 
therefore be useful to vary the dust law from Calzetti to SMC to Milky Way dust 
and to report the amplitude of uncertainties caused in other SED fitting parameters.  The idea of removing dust from the fit was discussed; while defensible for 
LAEs where an absence of dust could be assumed, this could make it more 
difficult to compare results with other galaxy types 
where dust is clearly present.  

{\bf 7. IGM absorption:}
A model for 
absorption by the intergalactic medium (IGM) must be applied to template 
spectra.  Most commonly this follows \citet{madau95} but during the discussion 
updated analyses by \citet{songaila04} and \citet{fauchergiguereetal08}
were recommended.  
These references provide a method for modeling IGM absorption of 
continuum emission, but the prescription that should be applied to Ly$\alpha$ 
emission is unclear.  Naively, all photons blue-wards of 1215.67{\AA} rest-frame 
suffer IGM absorption, but the treatment in the literature varies from 
assuming no IGM effect on Ly$\alpha$ to assuming that fully 
half the Ly$\alpha$ 
photons have been absorbed by the IGM.  Radiative transfer effects from 
resonant scattering and galactic winds are too complex to yield an obvious 
recipe, as the simulations of \citet{verhammeetal06} have shown.  This provides 
a motivation for removing Ly$\alpha$ emission from the SED fit by ignoring 
the narrow-band photometry, although more advanced options are described below.   

{\bf 8. Number of stellar populations:}
Fitting a simple stellar population to an LAE is not a good approach if one is 
interested in determining age and total stellar mass.  The presence of 
a starburst 
guarantees a young age and a low implied stellar mass for an SSP.  
Even CSF and $\tau$ 
models are poorly suited for modeling a star formation history that is likely 
to have made a rapid upward jump at the beginning of the starburst (possibly 
due to a galaxy merger).  Allowing exponentially increasing SFR through 
negative values of $\tau$ will help somewhat.  However, we need to utilize 
multiple-population SED models, as done by \citet{pirzkaletal07} and 
\citet{gawiseretal07}, to reveal any underlying population of old stars whose 
star formation history does not smoothly tie onto the active starburst.  

However, as noted above, the low S/N available with LAE SEDs makes it important to avoid 
fitting too many parameters.  Even an SED with 13 bands ($UBVRIzJHK$[3.6][4.5][5.8][8.0]) can lead to degeneracies when fitting a dozen parameters plus the overall normalization.  Each stellar population comes with 4 degrees of freedom:  instantaneous 
SFR, age, $\tau$, and $A_V$ (stellar mass is determined by the first three).
So we clearly cannot afford to fit more than two populations, and there are benefits to reducing the degrees of freedom by simplifying the populations e.g., 
by assuming a model at the age of the universe for the old population.  
\citet{gawiseretal07} assumed 
that both populations see the same dust extinction;  although 
this assumption is probably not true in general, it is very difficult to 
determine the dust reddening for an older population detected in only a couple 
IRAC bands where it dominates the photometry.  
It may make the most sense to combine a young model from Starburst99 with 
an old model from e.g. M05.  

{\bf 9. Individual SEDs or stacked?}
Low S/N makes it difficult to obtain robust results from SED fitting of 
individual LAEs, although several papers have chosen this approach.  
An alternative is to average (``stack'') the photometry of an 
entire sample of LAEs.  The latter approach achieves greater S/N at the 
cost of revealing ``average'' SED parameters for the entire LAE population 
rather than individual objects.  \citet{finkelsteinetal09} 
compared these methods and 
demonstrated that the average of individual LAE SED parameters was 
somewhat similar to 
the SED parameters fit to the average LAE SED, although a more detailed 
investigation is needed to be able to properly compare stacked and individual 
SED fit parameters.
  The discussion pointed out that if an analysis 
splits a population of objects into subsets based upon, e.g., [3.6]$\mu$m flux, one should perform simulations to identify and subtract Malmquist-type biases 
caused by noise in the resulting SED parameters (noise causes the [3.6]$\mu$m 
flux to be overestimated for the brighter stack and underestimated 
for the dimmer stack).  


{\bf 10. Include nebular emission lines?}
An important decision in SED fitting is whether to include nebular emission 
lines in the template spectra. 
 As \citet{zackrissonetal08} have pointed out,
these nebular emission lines can make a significant difference in SED 
fitting and photometric redshifts but are usually neglected.  
 For very young populations at $4<z<5$, H$\alpha$ 
can make a significant contribution to the [3.6]$\mu$m photometry.  
BC03 models report the number of ionizing photons, which can be turned 
into Ly$\alpha$ and Balmer series luminosities assuming Case B recombination.  
Starburst99 does this automatically and includes the 
critical [O II] and [O III] emission lines.

{\bf 11. Treatment of narrow-band photometry:}
Since spectral templates typically do not include emission lines and 
Ly$\alpha$ is particularly complicated, the most common approach has been 
to ignore the narrow-band photometry of LAEs.  Sometimes the inferred 
Ly$\alpha$ flux is used to subtract the emission-line contribution from 
overlapping broad-band photometry.  This is sensible, although the uncertainty 
in this correction should be propagated into the photometric uncertainties 
on the broadband fluxes.  
One caveat when using photometry from a narrow-band 
filter with a rounded (rather than top-hat) response curve is to avoid 
assuming that the LAEs are all at the redshift corresponding to the 
peak filter transmission, as this will systematically underestimate the 
objects' true Ly$\alpha$ fluxes.  Ideally the redshift should be varied, 
and the filter transmission can at least be represented by a more typical 
value \citep[see][]{gronwalletal07}.

A well-motivated 
attempt has recently been made by 
\citet{finkelsteinetal07,finkelsteinetal08a,finkelsteinetal09} 
to incorporate Ly$\alpha$ emission in the spectral templates to be compared with 
the full LAE SEDs including narrow-band photometry.  These 
authors introduced a new SED parameter, $q$, where $q=1$ implies trivial 
radiative transfer, $q<1$ implies the expected 
preferential extinction of Ly$\alpha$ photons, 
and $q>1$ implies that Ly$\alpha$ photons are enhanced versus continuum 
photons (either by anisotropic radiative transfer or possible clumpy dust 
as described by \citealp{neufeld91}).  
The combination of SFR, stellar age, and dust is sufficient to predict 
the Ly$\alpha$ flux when $q=1$, 
and comparison with the continuum-subtracted 
narrow-band flux density determines $q$.  The discussion appeared 
to produce agreement that in this sense $q$ is independent of the 
broad-band SED fit and could be determined 
subsequently to simplify computation.  

{\bf 12. Treatment of photometric uncertainties:}
  When population-averaged 
fluxes are being used, a bootstrap analysis can be performed to include 
sample variance in the photometric uncertainties; this usually dominates 
the formal uncertainty in the average flux so is important to include.  
The discussion illuminated concerns about systematic errors in photometry, 
due primarily to the difficulty of performing aperture photometry in the 
IRAC bands given the much larger PSF and significant source confusion.  

Some statistically dubious 
habits have crept into the LAE SED fitting literature with 
authors excluding ``non-detections'' from the fits or only penalizing 
the $\chi^2$ when a template flux exceeds a $3\sigma$ upper limit.  This 
alters the $\chi^2$ statistic to {\it no longer follow a $\chi^2$ distribution}, 
making interpretation difficult.  Moreover, this modification 
is entirely unnecessary when SED fitting is done in flux (magnitudes should 
be avoided at all costs given the asymmetry of errors at low S/N).  For low 
S/N data, noise fluctuations can cause negative fluxes, and these fluctuations 
are properly handled by feeding $\chi^2$ the observed fluxes and their 
formal photometric uncertainties.  The related 
practice of only plotting photometry 
for bands with formal $3\sigma$ detections is less dangerous but equally 
hard to justify.

{\bf 13. Method for determining best-fit model:}
Producing a likelihood for a single template is straightforward, but 
over a large parameter space the brute-force approach of determining the 
likelihood for every possible template can be very computationally intensive.
Some authors are simplifying this using Markov-Chain Monte Carlo (MCMC), for 
which public routines exist.  MCMC tries to avoid getting stuck in 
local minima, but extensive testing is still recommended to be sure that the 
global optimum is being found successfully.  
Another decision to be made is frequentist versus Bayesian interpretation 
of this likelihood \citep[see][]{lupton93}.  A frequentist analysis 
will simply call 
the template with maximum likelihood the best fit, whereas a Bayesian 
analysis will choose prior probabilities on all of the parameters and 
multiply the likelihood by 
 these to report the best fit parameter set.  

{\bf 14. Method for determining parameter uncertainties:}
The final choice to be made is how to determine the uncertainties in 
the best-fit SED parameters.  The most common method is to use $\Delta \chi^2$
where the $\chi^2$ value is allowed to increase by the right amount to 
represent the 95\% confidence level for the given number of 
degrees of freedom, $n_{\mathrm{dof}}$.  
However, 
frequentist analysis actually recommends ruling out only those parameter values 
for which no template is a good fit to the data 
(despite considering all possible values of other parameters) using 
absolute values of $\chi^2$ rather than $\Delta \chi^2$.  Unless 
the error bars are underestimated, this is 
quite conservative. 
Many astrophysicists instead use the Bayesian 
approach, where a ``credible region'' on a given parameter is determined 
by projecting the likelihood function onto that single dimension, marginalizing 
over all other parameters, and keeping the parameter range that contains 
e.g. 95\% of the posterior probability. To fully grasp parameter dependencies and correlations, the full dimensional parameter space must be studied.   
Monte Carlo simulations can also 
be performed by varying the observed photometry within its reported 
uncertainties and determining the range of best-fit parameters.  This is 
a reasonable approximation but is not statistically self-consistent.  
For example, a particular model might be an acceptable fit to all of these simulations 
but never end up as the best fit, so the parameter range of ``best fits'' 
can underestimate the true uncertainties.  MCMC codes produce uncertainties 
along with their best fits.  

Discussion ensued over how to determine the number of degrees of freedom.
It depends on whether a single model or an entire parameter space is 
being evaluated.  For a model, 
$n_{\mathrm{dof}} = n_{\mathrm{data}} - n_{\mathrm{nuisance}}$, where the 
final term is the number of nuisance parameters being fit such as 
the overall normalization.    
For a parameter space, 
$n_{\mathrm{dof}} = n_{\mathrm{data}} - n_{\mathrm{parameter}}$, which 
essentially corresponds to treating all parameters as nuisance parameters 
to see if {\it any} model in this entire space is an acceptable fit to the data.    

\section{Acknowledgments}
I thank the conference organizers for a well-organized, extremely pedagogically useful meeting, and participants in the SED fitting discussion for contributing 
ideas recounted here.  Len Cowie deserves special mention for making 
numerous valuable suggestions.  
I wish to thank Kim Nilsson for creating the 
figures used for my review discussion
at the meeting, Suraphong Yuma for providing the summary of 
LAE SED fitting results from 
his poster at the meeting and Steve Finkelstein for 
sending the error bars on his published SED fitting results shown in 
the table.  
I acknowledge 
financial support for LAE SED research from Spitzer Space Telescope 
archival programs AR-40823 and AR-50805.  




%
%
%

\end{document}